\documentclass[%
aip,
amsmath,amssymb,
reprint,%
]{revtex4-2}

\usepackage{amsmath} 
\usepackage[dvipsnames]{xcolor} 
\usepackage{amsmath}
\usepackage{amssymb}
\usepackage{epsfig}
\usepackage{tikz}
\usepackage{dsfont}

\usepackage[version=3]{mhchem} 

\begin{document}

\title{Scanning spin probe based on magnonic vortex quantum cavities}

\author{Carlos A. Gonzalez-Gutierrez}
\affiliation{Instituto de Nanociencia y Materiales de Arag\'on (INMA), CSIC-Universidad de Zaragoza, Zaragoza, ES-50009 Spain}
\affiliation{Department of Physics and Applied Physics, University of Massachusetts, Lowell, MA 01854 USA}
\affiliation{Instituto de Ciencias F\'isicas, Universidad Nacional Aut\'onoma de M\'exico, Av. Universidad s/n, Cuernavaca, Morelos, 62210, M\'exico}
\author{David Garc\'ia-Pons}
\affiliation{Instituto de Nanociencia y Materiales de Arag\'on (INMA), CSIC-Universidad de Zaragoza, Zaragoza, ES-50009 Spain}
\author{David Zueco}
\email{dzueco@unizar.es}
\affiliation{Instituto de Nanociencia y Materiales de Arag\'on (INMA), CSIC-Universidad de Zaragoza, Zaragoza, ES-50009 Spain}
\author{Mar\'ia Jos\'e Mart\'inez-P\'erez}
\email{pemar@unizar.es}
\affiliation{Instituto de Nanociencia y Materiales de Arag\'on (INMA), CSIC-Universidad de Zaragoza, Zaragoza, ES-50009 Spain}

\date{today}
\begin{abstract}
 Performing nanoscale scanning electron paramagnetic resonance (EPR) requires three essential ingredients. First, a static magnetic field together to field gradients to Zeeman split the electronic energy levels with spatial resolution. Second, a radiofrequency (rf) magnetic field capable of inducing spin transitions. Finally, a sensitive detection
 method to quantify the energy absorbed by spins. This is usually achieved by combining externally applied magnetic fields with inductive coils or cavities, fluorescent defects or scanning probes. Here, we {\color{black} theoretically propose the realization of a EPR scanning sensor merging all three characteristics into a single device}: the vortex core stabilized in ferromagnetic thin-film discs. On one hand, the vortex ground state generates a significant static magnetic field and field gradients. On the other hand, the precessional motion of the vortex core around its equilibrium position produces a circularly polarized oscillating magnetic field, which is enough to produce spin transitions.  Finally, the spin-magnon coupling broadens the vortex gyrotropic frequency, {\color{black} suggesting} a direct measure of the presence of unpaired electrons. Moreover, the vortex core can be  displaced by simply using external magnetic fields of a few mT, enabling EPR scanning microscopy with large spatial resolution. Our {\color{black} numerical} simulations show that, by using low damping magnets, it is {\color{black} theoretically} possible to detect single spins located on the disc's surface. Vortex nanocavities could also attain strong coupling to individual spin molecular qubits, with potential applications to mediate qubit-qubit interactions or to implement qubit readout protocols.
\end{abstract}

\maketitle

\section{Introduction}
Electron Paramagnetic Resonance (EPR) is widely used in chemistry, physics, medicine and material science to characterize the electronic structure of magnetic molecules and impurities\cite{Abragam2012}. This has important applications in the study of organic and inorganic free radicals,  colored centers in crystals, tissue oxygenation and archaeological dating, to give a few examples. Similarly to the physics of nuclear magnetic resonance imaging \cite{Levitt2008}, EPR can be combined with nanoscopic field gradients to detect spatially distributed spins \cite{Epel2017}.

The technological interest in EPR has led to the development of very sophisticated and sensitive detection methods. These include optical techniques, such as using optically active atomic defects  \cite{Simpson2017,Bucher2019,Healey2022,Klein2022}, electrical detection of magnetic signals using scanning tunneling microscopy probes \cite{Durkan2002,Berggren2016,Baumann2015} or mechanical sensing based on magnetic resonance force microscopy \cite{Rugar2004,Poggio2010}. Among the inductive methods, pickup coils and superconducting quantum interference devices (SQUIDs) have been employed to characterize resonant phenomena in small paramagnetic crystals \cite{Yue2017,Toida2016}. However, the most widespread inductive-EPR readout technique uses transmission lines and cavities \cite{Artzi2015, Blank2017, Abhyankar2022}. This latter approach offers the advantage of confining light in the space domain, yielding increased light-matter interactions. {\color{black} In addition,} the cavity's quality factor can be further enhanced when using superconducting coplanar resonators instead of metallic three dimensional cavities \cite{Goeppl2008}, {\color{black} yielding increased visibility}. This idea is exploited in circuit Quantum Electrodynamics (QED) \cite{Blais2007} , allowing, e.g., the manipulation and interrogation of superconducting or magnetic qubits \cite{Wallraff2004} and quantum sensing of small amounts of spins \cite{Kubo2012,Bienfait2015,Bienfait2016,Probst2017,Bienfait2017, Eichler2017,Ranjan2020,Hughes2021,Wang2023}.

The light-matter {\color{black} coupling factor (g)} depends on the amplitude of the (position-dependent) root mean square (rms) vacuum magnetic field fluctuations in the cavity $\textbf{B}_{\rm rms}(\textbf{r})$. 
The latter increases for decreasing mode volume, favouring the use of small cavities.
However, downsizing is limited by the maximum operating frequency ($1 - 20$ GHz, typically) and the impedance of the resonator. Large coupling factors can be reached by fabricating nanoscopic constrictions at the central transmission line in superconducting coplanar resonators \cite{Jenkins2014,Tosi2014,Haikka2017}. This approach keeps the total length of the cavity while reducing the other two dimensions, yielding strong focusing of the rms vacuum field fluctuations in nanometric regions around the central conductor.  By doing so, large coupling factors
of $g/2\pi \sim 1$ kHz per spin have been  demonstrated \cite{Gimeno2020}.  Alternatively, low impedance LC-resonators allow increasing the amplitude of the magnetic compared to the electric field fluctuations,  also yielding increased couplings \cite{Bienfait2015,Bienfait2016,Eichler2017,McKenzieSell2019,Wang2023}. 

In addition to photons, the solid state offers a wide variety of bosonic excitations that can be emitted or absorbed such as, e.g., quantized spin waves or magnons \cite{Graf2018,LachanceQuirion2019,Rameshti2022,Zheng2023}. {\color{black} In what follows, we will use $g$ to denote both the spin-photon and the spin-magnon coupling}. Magnonic cavities could be used to perform spin qubit readout or to mediate spin-spin interactions \cite{Trifunovic2013,Wolfe2014,Andrich2017,Flebus2019,Zou2020,Fukami2021,GonzalezBallestero2022,Karanikolas2022,Solanki2022,Xiong2022,Zou2022}, offering the advantage of increasing the coupling by operating at reduced wavelengths (compared to electromagnetic resonators of the same frequency). This is possible since spin wave modulation is only limited by the lattice constant of the ferromagnet, allowing the downsizing to the nanometer range. {\color{black} This significantly decreases the mode volume and results in substantial spin-magnon couplings.} For example, quasi-homogeneous spin waves in saturated ferromagnets have been proposed to  substitute superconducting cavities
\cite{Neuman2020,Wang2021a}. Using nanoscopic Yttrium-Iron-Garnet (YIG) spheres, such approach shall provide strong coupling to individual free electrons, even reaching $g/2\pi \sim 1$ MHz. 
Interestingly, whispering gallery modes in relatively large vanadium tetracyanoethylene (V[TCNE]$_x$) discs would yield a sizable spin-magnon coupling to individual NV centers \cite{Candido2020}. This is so for the relevant mode volume  is given by the disc's thickness $t$ and the angular index magnon mode. In this way, a very encouraging $g/2\pi \sim 10 $ kHz can be obtained with a $R =  500$ nm, $t = 100$ nm  out-of-plane saturated disc at $1.3$ GHz.
However, both approaches do require an externally applied bias field $B_{\rm ap}$ with a double purpose: to saturate the ferromagnetic volume and to tune its resonance frequency with that of the spins. 
Interestingly, magnons can be confined in peculiar flux-closure configurations at $B_{\rm ap}=0$. This is the case of magnetic vortices that are easily stabilized in thin-film ferromagnetic structures with lateral size between a few $10$ nm up to several micrometers \cite{Shinjo2000,MartinezPerez2020}. Minimization of magnetostatic energy yields a circular in-plane arrangement of spins with a nanoscopic out-of-plane magnetization core in the center. The vortex core modifies the spin-wave  spectrum of the ferromagnet, adding new resonant modes in the absence of externally applied fields \cite{Park2003,Novosad2005}.

Here, we compare the spin coupling to microwave resonators with that resulting from magnonic cavities (both saturated ferromagnets and flux-closure states).
In the case of magnons, we base our calculations on three archetypical materials, common in the field of quantum magnonics \cite{Haygood2021,Zhang2023,Schoen2016}: On the one hand,  a ferrimagnetic garnet having record low damping  but low saturation magnetization (YIG, with Gilbert damping parameter $\alpha \sim 10^{-4}$). On the other hand, two ferromagnets with larger magnetization but higher damping, i.e., Co$_{25}$Fe$_{75}$ alloy (CoFe, with $\alpha \sim 10^{-3}$) and Permalloy (Py, with $\alpha \sim 0.5 \times 10^{-2}$), the most widely used soft-ferromagnet for spin-wave-based applications. 
We first demonstrate that  the coupling to magnons is more than two orders of magnitude larger than that resulting from cavity photons. Although being of the same order, using vortices has important advantages over saturated ferromagnets such as the absence of a biasing magnetic field, independence from the saturation magnetization ($M_{\rm sat}$), and the capability to manipulate the vortex's position. This ability enables scanning over a range of tens of nanometers.
In summary, our findings highlight the potential of flux-closure states compared to other systems, demonstrating that the vortex core can be operated as a nanoscopic scanning EPR probe. Coupling the ferromagnetic disc to a superconducting circuit, our approach shall allow to spatially resolve the location of single spins distributed over the surface of the disc. 

\section{Results}


EPR is a resonant phenomenon, and regardless of the use of superconducting or ferromagnetic cavities, it must satisfy two conditions. 
%
The first criterion is energetic: the frequency \(\omega_0\) of the rf field produced by the cavity needs to match the energy difference between Zeeman-split spin levels. In the case of free \(S=1/2\) spins, the latter means \(\omega_s = \gamma_e B_{\mathrm{tot}}(\mathrm{r}) = \omega_0\). Here, \(\gamma_e/2\pi = 28\) GHz/T and \(B_{\mathrm{tot}}(\mathrm{r}) \equiv \left|\mathbf{B}_{\mathrm{tot}} (\mathrm{r})\right|\) is the total (position-dependent) magnetic field that contributes to the Zeeman splitting. 
 This criterion leads us to the definition of the \emph{resonance window}, i.e., the region in space where spins are resonant with the cavity (see  Section {\bf Methods}). We highlight that the resonance window will only depend on the modulus of the (position-dependent) $\mathbf{B}_{\mathrm{tot}}(\mathrm{r})$. 

The second condition is imposed by the geometry of the experiment: spin transitions can be induced only by the components of the vacuum field fluctuations that are perpendicular to the quantization axis of the spin. In the case of free \(S=1/2\) spins, the latter is parallel to \(\mathbf{B}_{\mathrm{tot}}\). This determines the strength of the spin-photon coupling $g$.
To quantify the coupling, it is convenient to introduce the set of mutually orthogonal vectors \(\mathbf{B}_{\mathrm{rms}, j}\), \(j=1,2,3\) so that \(\mathbf{B}_{\mathrm{rms},3} \parallel \mathbf{B}_{\mathrm{tot}}\). Thus, the other two components can induce spin transitions, allowing us to define \(B_{\mathrm{rms}}=\sqrt{B_{\mathrm{rms,1}}^2+B_{\mathrm{rms,2}}^2} \). Consequently,  in the case of \(S=1/2\), the spin-photon coupling is given by \(g = \mu_{\mathrm{B}} B_{\mathrm{rms}}\) with \(\mu_{\mathrm{B}}\) being the Bohr magneton.\footnote{This is a direct consequence of the choice of the orthogonal vectors \(\mathbf{B}_{\mathrm{rms}, j}\) and the formula \(g = g_e \mu_{\mathrm{B}} \langle 0 | \mathbf{S} \cdot \mathbf{B}_{\mathrm{rms}}(\mathbf{r}) | 1 \rangle\). Here, \(g_e\) is the gyromagnetic factor and \(\{|0\rangle, |1\rangle\}\) are the two states-induced spin transitions.} We highlight that $g$ will depend on both the intensity of the vacuum field fluctuations and also on the position-dependent distribution of its components with respect to $\mathbf{B}_{\mathrm{tot}}(\mathrm{r})$.  

\begin{figure*}
\includegraphics[width=0.9\textwidth]{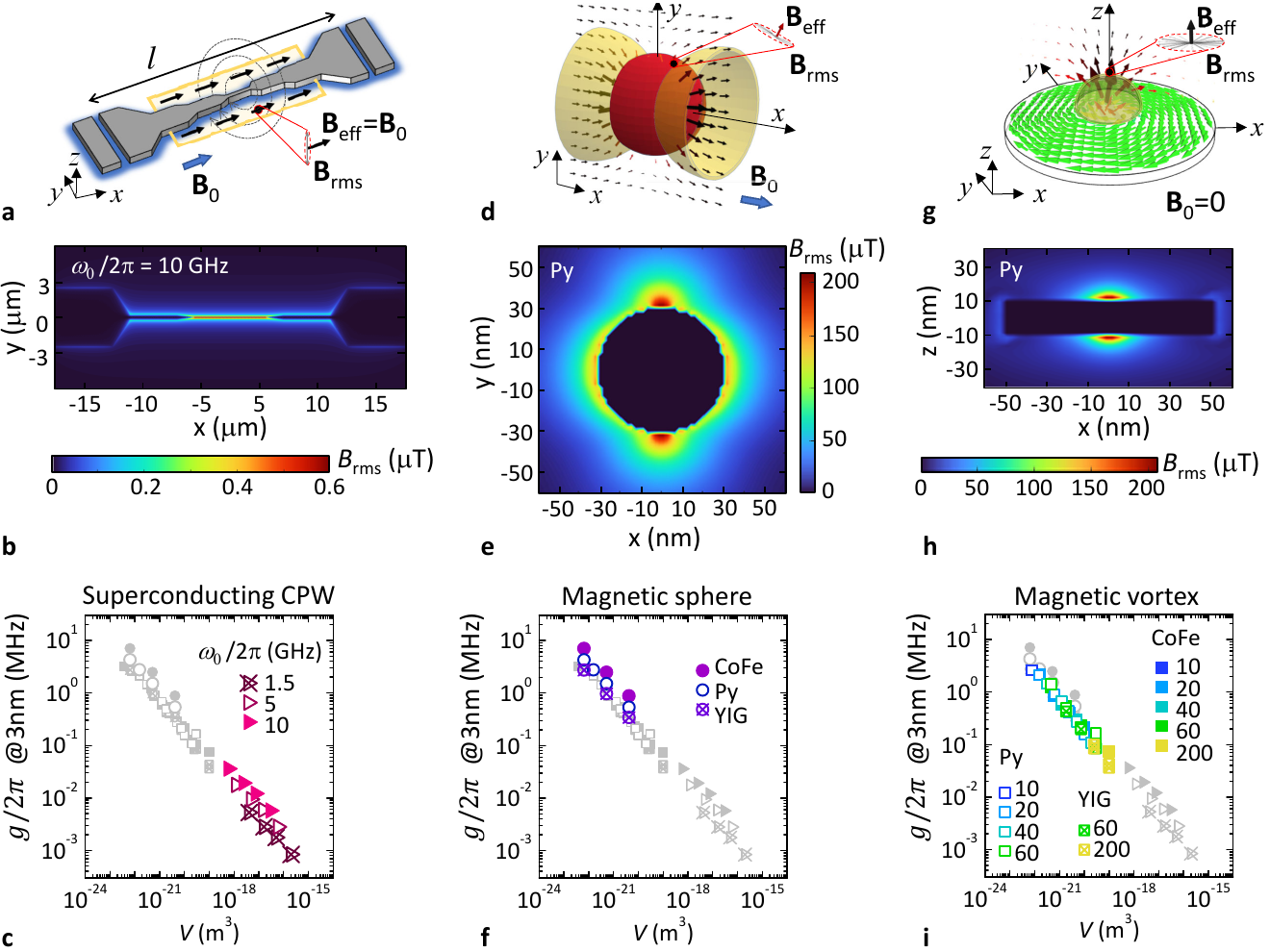}
\caption{
Comparison between electromagnetic cavities (coplanar waveguide) and magnonic resonators (saturated sphere and magnetic vortex). (a) (d) and (g): Scheme of the resonators.  $\textbf{B}_{\rm ap}$ is the external bias field  (big blue arrow). $\textbf{B}_{\rm tot}=\textbf{B}_{\rm ap} + \textbf{B}_{\rm stray}$ is the total field (black arrows), with $\textbf{B}_{\rm stray}$  the demagnetizing field and  $\textbf{B}_{\rm rms}$ the zero point fluctuation field  (gray arrows inside the red dashed circle). The yellow regions are the resonance windows.  Below we plot the spatial distribution of $B_{\rm rms}$ created by zero point current fluctuations in the electromagnetic cavity (b) and by the zero point magnetization fluctuations in a Py sphere (e) and a Py disc  (h) of similar volume. Bottom panels (c, f and i) are the calculated $B_{\rm rms}$ at $3$ nm above the resonators at the position of the white dot in panels b, e and h. Simulations evidence the expected dependence $B_{\rm rms} \propto V^{1/2}$.
}
\label{Fig1}
\end{figure*}

\subsection{Coupling to spins, comparison between photons and magnons}

We start by analyzing the spin-photon coupling in electromagnetic cavities. This will help us to compare with the case of magnons, which will be discussed below. We focus on the particular case of a superconducting coplanar waveguide formed by interrupting an open transmission line by two gap capacitors as sketched in Fig. \ref{Fig1}a. The external magnetic field \(\mathbf{B}_{\mathrm{ap}} = \mathbf{B}_{\mathrm{tot}}\) shall be ideally applied along the transmission line axis so that, in this case, \(B_{\mathrm{rms}} = |\mathbf{B}_{\mathrm{rms}}|\).
Under these circumstances, all spins can satisfy the resonance condition and are susceptible to be detected. 
In this case, the volume \(V\) of the resonator is the total length $l$ ({\color{black}see Fig. \ref{Fig1}a}) multiplied by the cross-section of the central transmission line. \(l\) is fixed by the operating frequency \(\omega_0\) which, together with the circuit's impedance \(Z_0\), sets the intensity of the zero-point current fluctuations \cite{Jenkins2013} \(i_{\mathrm{rms}} = \omega_0\sqrt{\hbar\pi/4Z_0}\).
Using electromagnetic waves imposes a lower limit on the total length of the cavity \(l > {\lambda}/{2} = {2\pi v}/{\omega_0}\), where \(\lambda\) and \(v\) are the wavelength and {\color{black} phase velocity}, respectively. Working at high frequencies allows increasing the coupling by reducing the volume, but this is limited to \({\omega_0}/{2\pi} < 10 - 15\) GHz due to technical reasons.
 Decreasing $V$ can only be accomplished by decreasing the cross-section which has the effect of focusing the vacuum field fluctuations while keeping $i_{\rm rms}$ unchanged. 
 Fig. \ref{Fig1}b shows the spatial distribution of $B_{\rm rms}$ for a $\omega_0/2\pi=10$ GHz resonator of thickness 50 nm and different widths ($5$ $\mu$m, $500$ nm and $50$ nm).   Fig. \ref{Fig1}c shows $B_{\rm rms}$ vs. $V$ calculated at 3 nm above the central conductor for different values of $\omega_0$ and the cross-section ($10 \times 10$ nm$^2$, $20 \times 20$ nm$^2$,  $35 \times 35$ nm$^2$ and $70 \times 70$ nm$^2$). From our simulations we see that the spin-photon coupling between free spins and $Z_0=50$ $\Omega$-resonators is limited to the range $1-50$ kHz, in agreement with recent experiments \cite{Gimeno2020}.


Alternatively, spin waves propagate at much lower velocities, allowing to  decrease $\lambda$ while working at frequencies in the 1 - 20 GHz range. This is the basic idea behind the use of magnonic cavities. One basic example to start with is the quasi-homogeneous Kittel mode  ($k = 2\pi/\lambda=0$) excited in isotropic ferromagnets, e.g., spheres.  
Zero point magnetization fluctuations in the ferromagnet produce zero point field fluctuations in the outside. 
Fig. \ref{Fig1}e shows the spatial distribution of $B_{\rm rms}$ for a saturated Py sphere for which  impressive values of 200 $\mu$T can be  reached. 
Fig. \ref{Fig1}f shows $B_{\rm rms}$   at 3 nm above the surface of spheres of different sizes ($R=25$, $50$ and $100$ nm) and materials (CoFe, Py and YIG).  
The intensity of field fluctuations does not depend on $B_{\rm ap}$ and {\color{black} satisfies} $B_{\rm rms} \propto V^{-1/2} M_{\rm sat}^{1/2}$, with $V$ the volume of the ferromagnet. 
As it can be seen, the resulting couplings $g = \mu_{\rm B} B_{\rm rms}$ will be more than two orders of magnitude larger than those achievable with superconducting resonators, even approaching the 10 MHz range. 

As a drawback, this approach requires unavoidably the use of an external bias field $\textbf{B}_{\rm ap}$ that serves to saturate the ferromagnet and to tune its resonance frequency $\omega_{\rm K}= \gamma_e B_{\rm ap}$. 
In addition, unlike the case of a superconducting resonator, paramagnetic spins located close to the sphere's surface will feel a strongly non-homogeneous magnetic field $\textbf{B}_{\rm tot} = \textbf{B}_{\rm stray} + \textbf{B}_{\rm ap}$ with $\textbf{B}_{\rm stray}$ the demagnetizing field created by the sphere itself (see arrows in Fig. \ref{Fig1}d).  The latter means that not all spins will satisfy the resonance condition $\omega_s = \gamma_e B_{\rm tot}=\omega_{\rm K}$. As a matter of fact, spins located at the position of the white dot in Fig. \ref{Fig1}e (where the coupling is maximum) will be far from resonance, therefore not contributing to the total signal.
Independently of the magnitude of $B_{\rm ap}$, the resonance window for free paramagnetic spins will be fixed and it corresponds to the yellow volumes shown in Fig. \ref{Fig1}d.  



Much more interesting for applications is the study of flux-closure topological magnetic textures like the magnetic vortex  sketched in Fig. \ref{Fig1}g. 
Without requiring the application of external magnetic fields, vortex dynamics include the gyrotropic precession of the vortex core around its equilibrium position at frequencies in the range $\omega_{\rm v0} \sim 0.1 - 2$ GHz. As an example,  Fig. \ref{Fig1}h shows $B_{\rm rms}$ created by a $R=50$ nm, $t=20$ nm Py disc. We obtain very large $B_{\rm rms}$ close to $200$ $\mu$T as in the  case of the Py sphere of similar volume (see Fig. \ref{Fig1}e). Finally, we compute $B_{\rm rms}$ at 3 nm from the disc surface. Results obtained for discs of different sizes ($R=50$, $100$ and $400$ nm and thicknesses given in the legend) and materials (CoFe, Py and YIG) are shown in Fig. \ref{Fig1}i. The intensity of the zero point field fluctuations is comparable to that obtained with the saturated spheres, also following the expected $V^{-1/2}$ dependence. Unlike the previous case, $B_{\rm rms}$ does not noticeably depend on $M_{\rm sat}$. 
In the following, we will analyze three distinct features that make vortex modes more suitable for spin detection compared to homogeneous modes.

\subsubsection{i) Absence of biasing field: the resonance window}


Vortex modes do resonate in the absence of any external biasing field  $B_{\rm ap} = 0$, which is a great advantage compared to saturated ferromagnets.  Besides, the demagnetizing stray field created by the vortex core itself reaches relatively large values close to the disc's surface (see dark arrows in Fig. \ref{Fig1}g).
 This stray field is enough to make the vortex gyrotropic mode resonant with paramagnetic $S=1/2$ spins fitting within the resonance window. In the case of the vortex, the latter takes the shape of a hollow semi-sphere (highlighted in yellow in Fig. \ref{Fig1}g). 
 The total effective coupling $G$ will be enhanced by a factor $\sqrt{N}$ where $N$ are the number of spins within the resonance window (see Section {\bf Implementation of the vortex sensor}). 

\begin{figure*}
\includegraphics[width=0.8\textwidth]{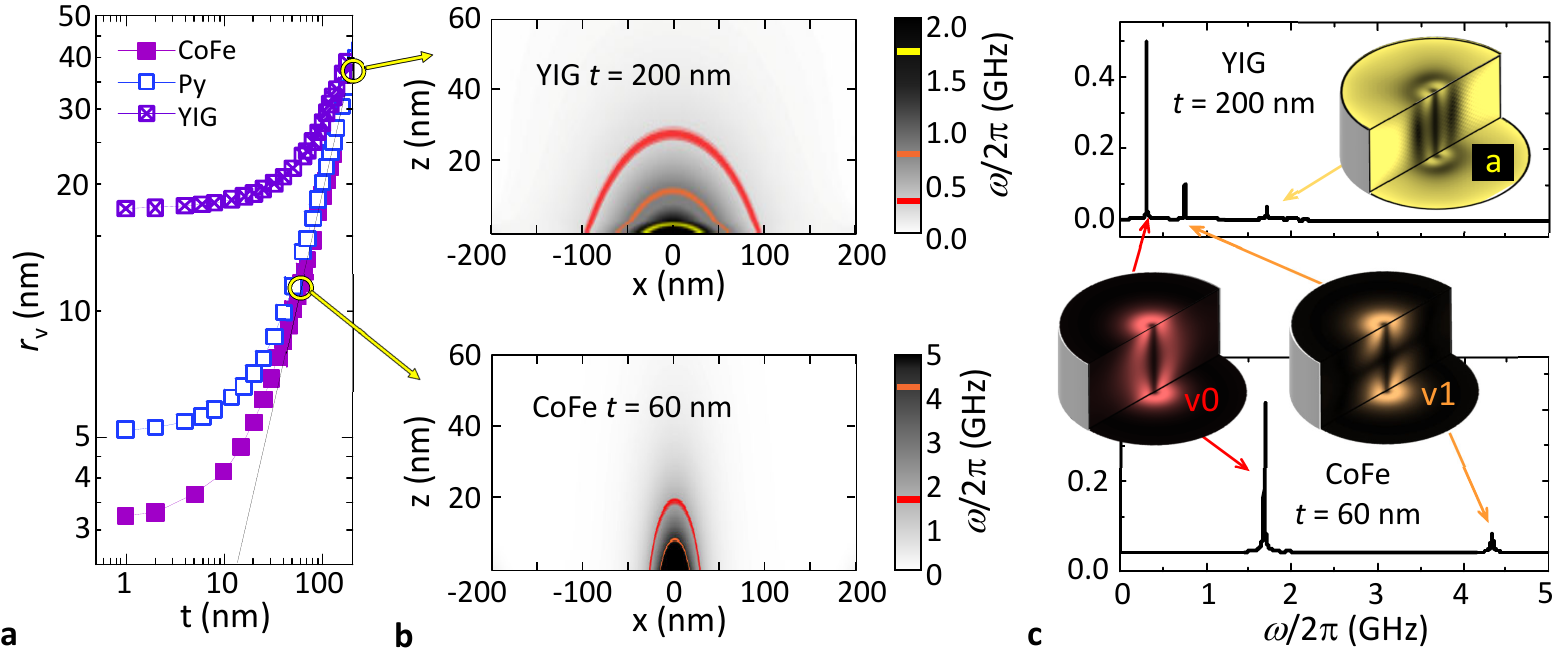}
\caption{(a) radius of the vortex core $r_{\rm v}$ vs $t$ for $R=200$ nm discs of different materials. $r_{\rm v}$ is nearly independent of $R$. (b) Spatial distribution of the resonance frequency for free $S=1/2$ spins for a YIG (top) and CoFe (bottom) discs with dimensions indicated. The latter is calculated as $\omega_s = \gamma_e B_{\rm demag}$. (c) Frequency spectrum of the same YIG (top) and CoFe (bottom) discs. The insets are the spatial  FFT of the out-of-plane time-dependent magnetization of the different modes. 
}
\label{Fig2}
\end{figure*}

The size of the resonance window depends on the radius of the vortex core itself  $r_{\rm v}$. The latter is given by the material-dependent exchange length $r_{\rm v} \simeq \lambda_{\rm ex} = \sqrt{ 2A / \mu_0 M_{\rm sat}^2}$ with $A$ the exchange stiffness and $\mu_0$ the vacuum magnetic permeability.  $r_{\rm v}$ is nearly constant, independent of the disc radius or thickness. However, for $t \gg \lambda_{\rm ex}$,  $r_{\rm v}$ increases linearly with the thickness (see Fig. \ref{Fig2}a). As a result, the resonance window increases considerably for materials with low $M_{\rm sat}$ and large thickness. This can be seen in Fig. \ref{Fig2}b, where we plot the spatial dependence of the spin resonance frequency $\omega_s = \gamma_e B_{\rm tot}$ for a paramagentic $S=1/2$ spin as a function of its position above a $R=200$ nm, $t=200$ nm YIG disc and a $R=200$ nm, $t=60$ nm CoFe disc. The resonance window corresponding to the fundamental gyrotropic mode of the two discs is highlighted in red, i.e., the region for which $\omega_s = \omega_{\rm v0}$. As it can be seen, the resonance window of YIG is considerably larger (diameter $\sim 190$ nm  and height $\sim28$ nm) than that of  CoFe ($\sim 54$ nm $\times 19$ nm). This stems from the (more than one order of magnitude) smaller $M_{\rm sat}$ of YIG compared to CoFe and the larger thickness of the YIG disc.


\subsubsection{ii) Independence on $M_{\rm sat}$ and higher order modes}

The vortex core gyro,  key to produce the $B_{\rm rms}$, {\color{black} also yields flexure oscillations of the vortex core line across the disc thickness \cite{Ding2014,Ding2014a}. For sufficiently thick discs, the latter leads to the emergence of higher order modes.} This is shown in Fig. \ref{Fig2}c where we plot the frequency spectrum of the same discs shown in panel (b).  Apart from the gyrotropic fundamental mode (denoted v0), we find a first flexure mode (v1). At higher energies it is also easy to find azimuthal modes (a) that correspond to the rotation of two halves of the disc with opposite out-of-plane net magnetization.  
These modes can be distinguished in the the FFT of the spatially-resolved time-varying out-of-plane magnetization in  the upper and bottom surfaces and the transverse cut of the discs, as shown in the inset of Fig. \ref{Fig2}c. Colored regions correspond to time-varying out-of-plane magnetization whereas darker areas correspond to constant magnetization. The signatures of vortex core precession can be seen in all modes. 

Materials with low saturation magnetization are very interesting for sensing applications as lowest dampings are usually obtained in insulating magnets like YIG or V[TCNE]$_x$. In this regard, the vortex mode offers an additional advantage over homogeneous Kittel modes for which $g \propto \sqrt{M_{\rm sat}}$. As shown in Fig. \ref{Fig1}i, spin coupling to vortex modes is nearly independent of $M_{\rm sat}$ encouraging the use of ultra low-damping ferrimagnets. However, low $M_{\rm sat}$ yields vortex gyration at sub-GHz frequencies, usually too low for practical readout circuits. In this way, flexure or azimuthal modes shall allow operating ultra-low damping ferrimagnets at reasonable frequencies. As an example, see the upper panel in Fig. \ref{Fig2}b where we show the resonance windows (orange and yellow) for high frequency modes in a $R=200$ nm, $t=200$ nm YIG disc ($\omega_{\rm v1 }/2\pi = 760$ MHz and $\omega_{\rm a}/2\pi = 1.7$ GHz, respectively). Interestingly, the resulting spin-magnon coupling to high frequency modes is comparable to that of the fundamental mode. To illustrate this, we calculate the coupling for a single $S=1/2$ spin located at the center, at 3 nm above the surface of the YIG disc. This yields $g_{\rm v0}/2\pi = 100$ kHz, $g_{\rm v1}/2\pi = 99$ kHz and $g_{\rm va}/2\pi = 84$ kHz for the fundamental, first flexure and first azimuthal modes, respectively.  


\begin{figure}[h]
\includegraphics[width=0.99\columnwidth]{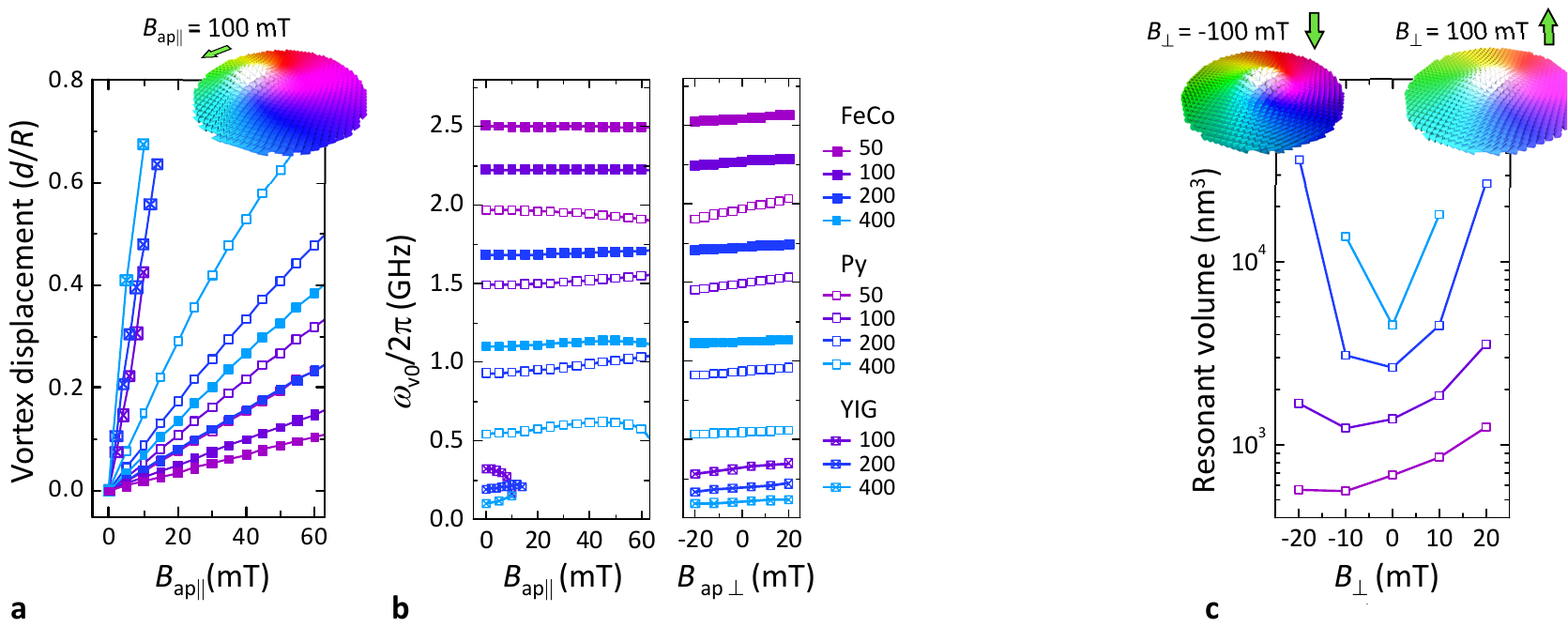}
\caption
{ 
(a) Normalized displacement of the vortex core $d$ upon an external in-plane bias field $B_{\rm app\parallel}$ and (inset) spatial distribution of the magnetization of a $R=50$ nm Py disc under $B_{\rm app\parallel}=100$ mT. (b) Resonance frequency for the v0 mode vs the external in-plane $B_{\rm app\parallel}$ (left) and out-of-plane field $B_{\rm app\perp}$ (right). 
In all panels, simulations are made for $t=60$ nm and different radius (in nm) and materials as indicated in the legend.
}
\label{Fig3}
\end{figure}

Additionally, high frequency modes in high $M_{\rm sat}$ materials would resonate with paramagnetic spins lying much closer to the disc surface, where the coupling is larger. This reduces the total volume of the resonance window, boosting the spatial resolution of the vortex sensing probe. For example, in the case of the $R=200$ nm, $t=60$ nm CoFe disc, the resonance window reduces from $\sim 54$ nm $\times 19$ nm (mode v0) down to $\sim 31$ nm $\times 9$ nm (mode v1).



\subsubsection{iii) Vortex mobility: Scanning spin detector}

 One of the most interesting properties of a spin sensor is the ability to scan over the surface of the sample. This is not trivial in the case of saturated ferromagnets but can be easily done in the case of magnetic vortices by using in-plane magnetic fields. The effect of an in-plane external field $B_{{\rm app}\parallel}$ is to enlarge the disc region having magnetization pointing in the same direction  while decreasing the size of the region having opposite magnetization (see inset in Fig. \ref{Fig3}a). This results in an effective translation of the vortex core position $d$ perpendicular to the direction of $B_{{\rm app} \parallel}$. Fig. \ref{Fig3}a shows the numerically calculated values of $d$ vs. $B_{{\rm app} \parallel}$ for 60 nm-thick discs of different materials and radius. The vortex core moves progressively as $B_{{\rm app} \parallel}$ increases until it approaches the edge of the disc where it is annihilated. 
{\color{black} We highlight that the vortex state remains stable up to the annihilation field. Experimental measurements indeed demonstrate that the energy barrier for vortex annihilation remain substantial, reaching several hundred K, even at large magnetic fields of several tens of mT \cite{MartinezPerez2020}. The latter holds true for discs of large radius but also for very small size, even below $R = 100$ nm. Additionally,} the displacement of the vortex has little effect on the gyrotropic frequency as shown in Fig. \ref{Fig3}b (left){\color{black}\cite{Aliev2009}}.  
The coupling strength and the size of the resonance window does not vary notably as long as the vortex displacement $d < R/2$. For larger displacements, the resonance window is considerably enlarged so that vortex fluctuations still allow spin detection.
The resonance can be also modified by means of an out-of-plane magnetic field $B_{\rm app\perp}$ as shown in Fig. \ref{Fig3}b (right){\color{black}\cite{Loubens2009}}. 
Interestingly, $B_{\rm app\perp}$ has also the effect of increasing (decreasing) the total size of the vortex core when applied parallel (anti-parallel) to the vortex polarity. Combining these two effects, it is possible to tune the total size of the resonance window. 



\subsection{Implementation of the vortex sensor}

\begin{figure}[h]
\includegraphics[width=0.99\columnwidth]{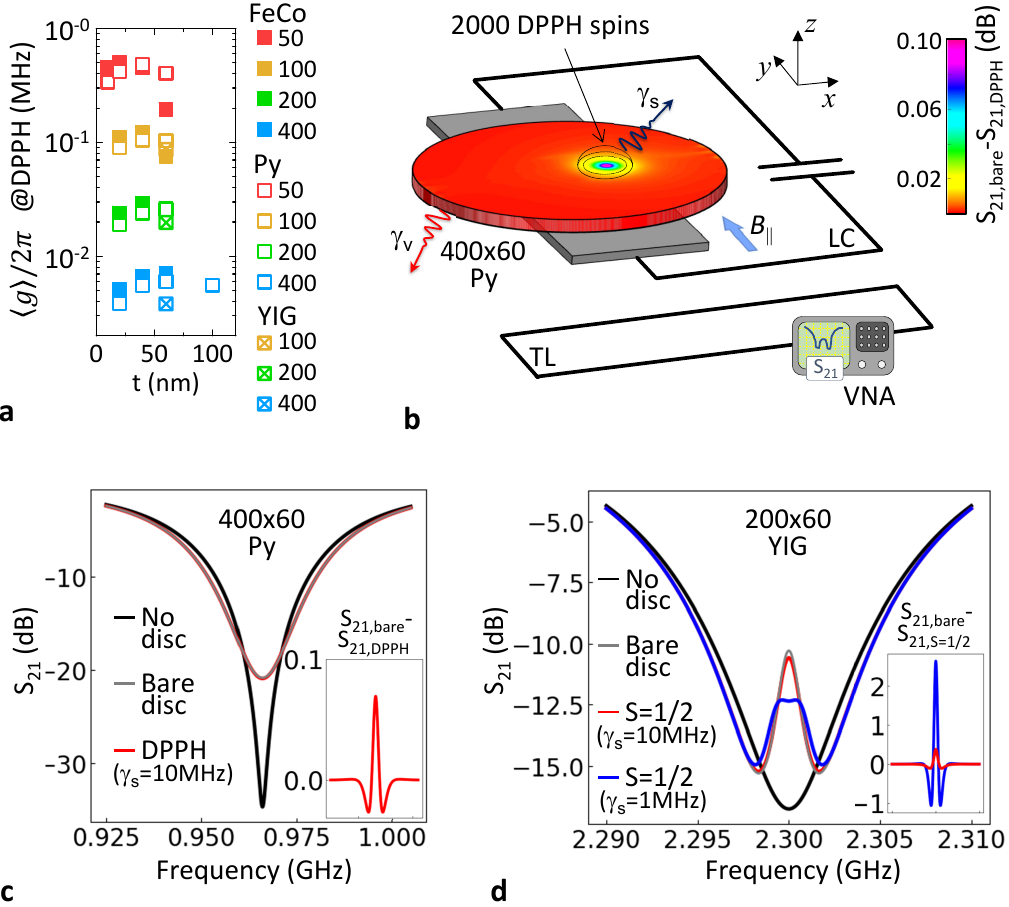}
\caption{(a)  $\langle g \rangle$ vs $t$ for different materials and radius (nm)  indicated in the legend. (b) A Py disc is located over a superconducting LC resonator coupled to a transmission line (TL) and the $S_{21}$ parameter is measured. The density plot shows the difference between the transmission at resonance with and without a small drop of DPPH at $d=52$ nm from the disc center. 
(c) Calculated  $S_{21}$ with no disc, with a bare Py disc and for a disc containing a drop of $\sim 2000$ DPPH spins. The inset shows $S_{\rm 21,bare}-S_{\rm 21,DPPH}$ in  dB {\color{black}with the horizontal axis and scale being the same as in the main panel}.  (d) Transmission for a YIG disc (under identical conditions as in panel a)  with $B_{\rm app\perp}=5$ mT. $S_{21}$ is calculated with no disc, with a bare YIG disc and with a disc containing a single spin $S=1/2$ at 3 nm over the surface with linewidth $\gamma_s = 10$ MHz and 1 Mz. The inset shows $S_{\rm 21,bare}-S_{\rm 21,S=1/2}$ in dB {\color{black}with the horizontal axis and scale being the same as in the main panel}.
}
\label{Fig4}
\end{figure}

To finish, we discuss  the experimental use of magnetic vortices to perform nanoscopic scanning imaging of a spin ensemble. The total spin-magnon coupling can be calculated as $G =\sqrt{\sum_j g_j^2}$, where $j$ goes from 1 to $N$ spins satisfying the resonance condition. 
The sensitivity relies on the ratio between $G$ and both the losses on the vortex resonator ($\gamma_{\rm v}$) and the linewidth of paramagnetic spins ($\gamma_s$). Increasing the coupling is achieved by using discs of small volume since $g \propto V^{-1/2}$ (cf. Fig. \ref{Fig1}i). However, the smaller the disc the more difficult its fabrication, manipulation and readout will be, so we will keep $R \sim 100 - 200$ nm. The average coupling per spin defined as $\langle g \rangle = \sqrt{\sum_j g_j^2 /N}$ does not depend noticeably on the disc thickness (see Fig. \ref{Fig4}a). For this reason, we will keep relatively large $t \sim 60$ nm, favoring the stabilization of vortices with not too small resonance frequencies.

If the total effective spin-magnon coupling satisfies $G>\gamma_{\rm v}, \gamma_s$, we are in the strong coupling regime and the resonance peak of the vortex mode splits into two peaks separated by $2G$. Measuring the splitting provides, therefore, a direct way of detecting the presence of spins as, e.g., the vortex core is scanned. On the other hand, sensing will be also possible in the weak coupling regime. In this case, the width of the vortex resonance $\gamma_{\rm v}$ will be enlarged if enough paramagnetic spins in the surface of the nanodisc satisfy the resonance condition. The linewidth $\gamma_{\rm v}$ can be experimentally measured using superconducting microcircuits that can be optimally coupled to magnonic resonators \cite{Huebl2013, MartinezPerez2018,MartinezPerez2019,Hou2019,Li2019,Haygood2021,Golovchanskiy2021,Li2022,Rincon2023}. In this way, the target spins can be deposited in powder or crystal form or from solution on the surface of the disc for nanoscopic EPR imaging.

A suitable experimental approach is depicted in Fig. \ref{Fig4}b. 
We consider a $R=200$ nm, $t=60$ nm Py disc that resonates at  $\omega_{\rm v0} /2 \pi =0.93$ GHz with losses $\gamma_{\rm v} = 14$ MHz. We chose Py although being a lossy ferromagnet as it is probably the most common and easy material to fabricate and manipulate. The disc is located on the inductive part of a superconducting LC resonator of width $100$ nm and thickness $50$ nm, in good contact to it. The LC resonator is interrogated through a superconducting transmission line. Under such circumstances, the absorption dip produced by the resonant disc shall produce changes in the transmission that amount to several tens dB (see Fig. \ref{Fig4}c). Now, we take the case of a typical EPR calibrating sample of free-radical molecules of 2-diphenyl-1-picrylhydrazyl (DPPH) having a density of $\rho \sim 2$ spins/nm$^2$, $S=1/2$, $g_e=2$ 
and $\gamma_s/2\pi \sim 10$ MHz. 
Scanning the vortex sensor, one could detect the presence of small drops containing $2\times 10^{3}$ DPPH spins (or, put in other words, volumes of only 0.4 atto-liter with 2 spins per nm$^2$), yielding $G/2 \pi=1.2$ MHz.  The signature of coupling would be a variation of $0.1$ dB of the resonance absorption as an external in-plane magnetic field is scanned (see Fig. \ref{Fig4}b and inset in panel c). 
Even more interesting for sensing applications, the use of low-damping materials like YIG opens the way to the detection of single spins. 
For instance, the first flexure mode in a $R=100$ nm $t=60$ nm YIG disc would bring one single spin (lying at 3 nm over the surface) into resonance at $\omega_{\rm v1} /2 \pi = 2.3$  GHz, yielding $G/2 \pi=0.7$ MHz. For this purpose, an out-of-plane biasing field of $5$ mT is necessary to tune the size of the resonance window. Under identical conditions as shown in Fig. \ref{Fig4}b and thanks to the low losses of YIG ($\gamma_{\rm v}=0.6$ MHz), this configuration would yield a large transmission signal depending on the relaxation time of the spin. This is exemplified in Fig. \ref{Fig4}d where the transmission coefficient $S_{21}$ resulting for a bare disc and a disc coupled to one individual spin with $\gamma_s/2\pi = 10$ MHz ($0.3$ dB difference) and  $1$ MHz ($3$ dB difference) are compared.

\section{Conclusions}

Our simulations suggest that, using Py, it is possible to detect small drops of 0.3 atto-liter containing 2 spins per nm$^2$ over a surface of $2 \pi \times 200^2$ nm$^2$. Additionally, 
the vortex core can be easily scanned by means of external magnetic fields
so to perform EPR scanning microscopy. Interestingly, the same can be in principle achieved using dissipation-free spin currents. In this way, vortex-based EPR microscopy could be implemented \emph{without} any externally applied magnetic field. High spin sensitivity stem from the small mode volume of the gyrotropic resonances {\color{black} which is independent of material parameters such as the saturation magnetization (cf. Fig. \ref{Fig1}i)}. This is an important advantage of the vortex probe sensor compared to saturated ferromagnetic nano-objects, encouraging the use of ultra-low damping ferrimagnets that usually come with reduced saturation magnetization. This property, together with the possibility of using high frequency vortex modes, makes it possible to reach very large spin-magnon couplings. For example, a $R=100$ nm YIG disc would allow detecting single spins with typical relaxation times of $\gamma_s \sim 10$ MHz. 


Our approach is also potentially very interesting to increase the weak interaction between superconducting microcircuits and  spin qubits, e.g., molecular qudits based on single rare earth ions that  offer transitions between tunnel-split ground state doublets with high spin \cite{GaitaArino2019}. Using low-damping magnetic vortices could make it possible to read the state of individual spin qudits, which was previously considered unattainable \cite{Jenkins2013}. Finally, we have also shown how to numerically normalize any magnon mode in ferromagnets of arbitrary size and shape. This can be used to calculate the zero-point magnetization fluctuations in confined nanomagnets including homogeneous magnon modes but also spin textures like domain walls, vortices or skyrmions. 

\section{Methods}

\subsection{Micromagnetic simulations}

We use the finite difference software MUMAX3 \cite{Vansteenkiste2014} to solve the time-dependent Laudau-Lifshitz-Gilbert equation for a given sample geometry and material. The relevant material parameters are the saturation magnetization $M_{\rm sat}$, the exchange stiffness $A$ and the Gilbert damping $\alpha$ (CoFe: $M_{\rm sat} = 1.9 \times 10^6$ A/m, $A = 2.6 \times 10^{-11}$ J/m and $\alpha =  10^{-3}$. Py: $M_{\rm sat} = 0.86 \times 10^6$ A/m, $A = 1.3 \times 10^{-11}$ J/m and $\alpha=  1 \times 10^{-2}$. YIG: $M_{\rm sat} = 0.14 \times 10^6$ A/m, $A = 0.37 \times 10^{-11}$ J/m and $\alpha=  10^{-5}$). In general, a disc of radius $R$ and thickness $t$ (or sphere of radius $R$) is simulated within a box with section $2R \times 2R$ and thickness large enough to calculate the relevant stray fields at 100 nm above the surface of the ferromagnet. The lateral size of the cells (with volume $v_{\rm cell}$) is kept below o approximately equal to $\lambda_{\rm ex} = \sqrt{ 2A / \mu_0 M_{\rm sat}^2}$.  

Magnon dynamics are simulated using a dc biasing field $\textbf{B}_{\rm ap}$ applied along a given direction, e.g. $\hat x$ in Fig. \ref{Fig1}d. $B_{\rm ap}$ is essential to obtain ferromagnetic resonances in the case of the saturated sphere  but is unnecessary when dealing with vortices that do also resonate at $B_{\rm ap}=0$. A sinusoidal time-dependent perturbation field $\textbf{B}=\beta {\rm sinc}  (\omega_{\rm cutoff} t)$ is applied perpendicularly to $\textbf{B}_{\rm ap}$ in the case of the sphere ($\hat y$ direction) and parallel to the disc plane ($x,y$) in the case of vortices. This is equivalent to exciting all spin-waves at frequencies below $\omega_{\rm cutoff}$. We identify the magnon modes by calculating the numerical FFT of the resulting time-dependent spatially-averaged magnetization (see, e.g., Fig. \ref{Fig2}c). We can write the dynamics at each cell for every mode, either a Kittel (K) or a vortex (v) one,  as
\begin{equation}
\label{m-methods}
    {\bf m}_\xi ({\bf r}_n, t)
 =
 {\bf A} ({\bf  r}_n)  e^{i \omega_\xi t} + {\rm c.c.} \qquad
 \xi = {\rm K}, \, {\rm v}
\end{equation}
with ${\bf r}_n$ the cell position, ${\bf A} ({\bf r}_n) $ the amplitude and $\omega_\xi$ their frequency.

To calculate the spin-magnon coupling we simulate the stray field, $\textbf{B}_{\rm rms} ({\bf r}_j)$, generated by the  
(zero-point-fluctuations of the) vortex magnetization from which the coupling can be obtained: 
\begin{equation}
\label{single-spin-coupling}
  g_j =\mu_{\rm B} B_{\rm rms} ({\bf r}_j)  \; .
\end{equation}
 For this purpose, we calculate the magnetic response on resonance using a perturbation field $B=\beta {\rm sin}  (\omega_{\xi} t)$. Here $\omega_{\xi}$ is the mode frequency and $\beta$ must be low enough to keep the system in the linear response regime, i.e., $\beta <$  200 - 500 nT. By doing so, we obtain, on the one side, the time-dependent vector magnetization at each cell $n$ inside the ferromagnet. 
 The amplitudes ${\bf A}$ in \eqref{m-methods} depend on the perturbation  strength  $\beta$.  Therefore, we must \emph{normalize} them to having the single-magnon amplitudes.  Using the usual formalism for magnon {\color{black} quantization} the magnetization vector is normalized as \cite{Mills2006}
\begin{align}
\Lambda (\beta)= \sqrt{ \frac{2g_e \mu_{\rm B} M_z}{v_{\rm cell} \sum_n A_x ({\bf r}_n) A_y({\bf r}_n)|\sin \Big (\delta_x ({\bf r}_n)- \delta_y ({\bf r}_n) \Big )|}}.
\label{norm}
\end{align}
Here, $M_z$ is the spatial-averaged component of the magnetization along $B_{\rm ap}$ in the case of the sphere and out-of-plane in the case of vortices.
$A_\alpha ({\bf r}_n)$ are the perpendicular (the in-plane for the vortex case)  amplitudes in \eqref{m-methods} while $\delta_x ({\bf r}_n)- \delta_y ({\bf r}_n)$ is the phase difference between these two components. On the other side,  we obtain the stray magnetic field resulting at every spin position ${\bf r}_j$ outside the ferromagnet $\textbf{B}_{\rm stray} ({\bf r}_j)=\textbf{B}^{\rm dc}_{\rm stray} ({\bf r}_j) + \textbf{B}^{\rm ac}_{\rm stray}(\beta, t; {\bf r}_j)$ where we have split the static and the time-dependent components and the dependencies on position, time and perturbation field $\beta$ are highlighted.  The total zero-point field fluctuations are given by $\textbf{B}_{\rm rms} ({\bf r}_j) = \Lambda \textbf{B}^{\rm ac}_{\rm stray} ({\bf r}_j) $, which is independent of $\beta$. Finally, we can calculate the contribution of $\textbf{B}_{\rm rms}$ able of inducing spin transitions. The latter is given by $B_{\rm rms} ({\bf r}_j)=\sqrt{B_{\rm rms,1}^2 + B_{\rm rms,2}^2}$ where $B_{\rm rms,1}$ and $B_{\rm rms,2}$ are the components of $\textbf{B}_{\rm rms}$ perpendicular to $\textbf{B}_{\rm tot} ({\bf r}_j) = \textbf{B}^{\rm dc}_{\rm stray} ({\bf r}_j) +  \textbf{B}_{\rm ap}$ (see Section {\bf Results}). 
Inserting $B_{\rm rms} ({\bf r}_j)$ into \eqref{single-spin-coupling}, the position-dependent spin-photon coupling can be obtained. {\color{black} We emphasize that Fig. \ref{Fig1}f and i show the spatial dependence of $B_{\rm rms} ({\bf r}_j)$, i.e., the part responsible for inducing spin transitions only.}

The resonance window is also calculated numerically. For this purpose we consider only those spins for which the energetic criterion is satisfied, i.e., $\omega_s = \gamma_e B_{\mathrm{tot}} = \omega_\xi$. This resonance window is enlarged by the broadening of both the magnonic resonance and the spins. In general, it will be reasonable to consider those spins that satisfy:
\begin{eqnarray}
\omega_\xi - \gamma_{\rm v}/2   \; \; \le  \; \; \gamma_e B_{{\rm tot,} j} \; \; \le  \; \; \omega_\xi + \gamma_{\rm v}/2,
\label{reswindow}
\end{eqnarray}
with $\gamma = \max(\gamma_{\rm v}, \gamma_s)$.

Finally, the average magnon-spin coupling per spin can be calculated as [Cf. Eq. \eqref{single-spin-coupling}]:
\begin{eqnarray}
\langle g \rangle = \frac{\sqrt{\sum_j g_j^2}}{\sqrt{N}},
\label{effcoup}
\end{eqnarray}
where the sum considers only those cells within the resonance window. This is to say, \eqref{effcoup} is the sum of the position dependent coupling divided by the square root of the total number of spins that satisfy the resonance condition.


\subsection{Electromagnetic simulations}

In the case of the superconducting resonators, the spatial distribution of $\textbf{B}_{\rm rms,s}$ is calculated using the finite-element code 3D-MLSI \cite{Khapaev2002}. This software  solves London equations in a superconducting circuit with given dimensions and London penetration depth $\lambda_{\rm L}$. In the simulations we set  $\lambda_{\rm L}=90$ nm for Nb and a flowing current $i_{\rm rms}$.  From here, 3D-MLSI allows calculating the spatial distribution of supercurrents and the resulting $\textbf{B}_{\rm rms,s} ({\bf r}_j)$. Under the particular geometry described in Fig. \ref{Fig1}a, all spins satisfy the resonance condition and $B_{\rm rms,s} ({\bf r}_j)=|\textbf{B}_{\rm rms,s} ({\bf r}_j)|$. The spatial-dependent coupling is given by Eq. \eqref{single-spin-coupling}.

\section{Supplementary Information}

\subsubsection{Theory for the vortex-sensor implementation: coupling between spins, vortex, LC-resonator and transmission line}
\begin{figure}
\includegraphics[width=0.99\columnwidth]{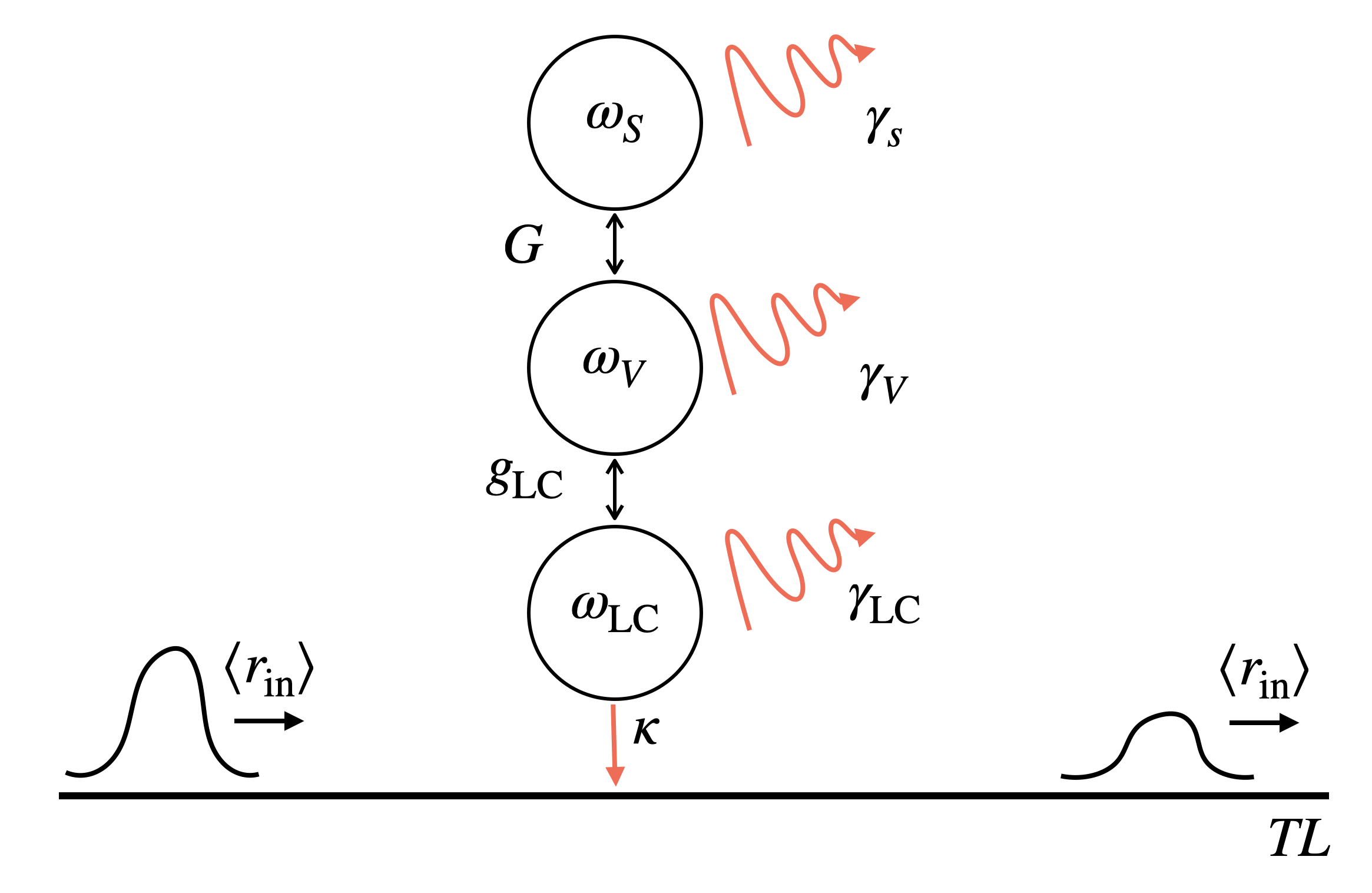}
\caption{Sketch of a transmission experiment.  The input signal is sent through a transmission line interacting with the LC-circuit.  In this sketch the modes (the LC-resonator, the vortex and the spins) are represented as circles, the couplings and dissipation channels are indicated.  The dynamics of the model is given in Eq. \eqref{eom}. }
\label{FigS1}
\end{figure}

The experimental setup for spin detection discussed in the main text involves a LC resonator coupled to a transmission line. Simultaneously, the disc is coupled to the LC resonator (mounted on top of the inductor), as shown in Fig. 4 in the main text. The stabilized vortex scans the spins.
The spin detection relies on the coupling of the latter to the vortex, resolved through a transmission experiment, that we model here. 

An input coherent signal is sent through the transmission line, and the output is measured after interaction with the complete system: LC resonator + disc + spins.
For convenience and clarity, we illustrate the coupling topology in Fig. \ref{FigS1}. We observe that the LC is directly coupled to the TL. Thus, employing input-output theory, this transmission is given by:
\begin{equation}
\label{s21}
S_{21} (\omega) = \frac{\langle r_{\rm out} \rangle}{\langle r_{\rm in} \rangle} = 1 - \kappa \chi_a (\omega) \; .
\end{equation}
Here, $\langle r_{\rm in, out} \rangle$ represents the averages of the input and output right-moving fields, $\kappa$ quantifies the coupling to the TL, and $\chi_a$ is the LC response, which can be computed as:
\begin{equation}
\label{chia}
\chi_a (\omega) = \frac{\langle a \rangle}{\sqrt{\kappa} \alpha_{\rm in}} \; .
\end{equation}
Here, $\langle r_{\rm in} \rangle = \alpha_{\rm in} e^{-i \omega t}$. Hence, according to the input-output theory, the transmission reduces to the calculation of the LC response function in this case.
To compute the LC response function, we employ a coupled mode model, depicted in Fig. \ref{FigS1}, where the LC, vortex, and spin are modeled as harmonic modes coupled among themselves. The dynamics are given by:
\begin{widetext}
\begin{align}
\label{eom}
\frac{d}{dt}
    \begin{pmatrix}
    \langle a \rangle
    \\
      \langle a_V \rangle
      \\
        \langle a_S \rangle
        \end{pmatrix}
        =
        \begin{pmatrix}
        -i \omega_{{\rm LC}} - \kappa -\gamma_{\rm LC} & -i g_{\rm LC} & 0
        \\
        - i g_{\rm LC} & -i \omega_V - \gamma_{\rm v} & -i G 
        \\
        0 & -i G & -i \omega_S - \gamma_S  
        \end{pmatrix}
         \begin{pmatrix}
    \langle a \rangle
    \\
      \langle a_V \rangle
      \\
        \langle a_S \rangle
        \end{pmatrix}
        +
        \begin{pmatrix}
        -i \sqrt{\kappa} \alpha_{\rm in} e^{-i \omega t}
        \\
        0
        \\
        0
        \end{pmatrix}
\end{align}
Here, $\langle a \rangle$, $\langle a_V \rangle$, and $\langle a_S \rangle$ represent the modes for the LC resonator, vortex, and spins, respectively. The remaining parameters include the LC-vortex coupling, $g_{\rm LC}$, the (collective) spin-vortex coupling, $G$, and the line-LC coupling, $\kappa$.  Finally, $\gamma_{\rm LC}$ denotes the intrinsic losses of the resonator. The vortex and spin dissipation, $\gamma_{\rm v}$ and $\gamma_s$, are also taken into account in the model.
Equation \eqref{eom} represents a linear set of differential equations that can be solved by moving to the interaction picture $\langle a \rangle \to \langle a \rangle e^{-i \omega t}$. This transformation yields an algebraic set of three coupled linear equations, which admit an analytical solution from which $\langle a \rangle$ can be obtained. Consequently, $S_{21}$ in Eqs. \eqref{s21} and \eqref{chia} is given by:
\begin{align}
\label{S21-final}
S_{21} (\omega) =
    1- 
    \frac{\kappa}
    {\kappa + \gamma_{\rm LC} + i (\omega_{\rm LC} - \omega)
    + g_{\rm LC}^2
    \frac{\gamma_S + i (\omega_S - \omega)}
    {G^2 + \Big ( \gamma_S + i (\omega_S - \omega) \Big ) 
    \Big ( \gamma_{\rm v} + i (\omega_V - \omega) \Big ) 
    }
    }
    \; .
\end{align}
\end{widetext}
The typical limiting cases of $G= g_{\rm LC} = 0$ (just the resonator) or the resonator plus vortex $G=0$ are easily recognized.

The resonance frequency of the superconducting cavity is given by $\omega_{\rm LC}=1/ \sqrt{LC}$ with $L$ and $C$ the inductive and capacitive components of the circuit, respectively. $L$ can be tuned by flux-biasing a superconducting quantum interference device (SQUID) coupled in series to the inductive part of the resonator \cite{Planat2019,Uhl2023}. In this way, $\omega_{\rm LC}$ can be adjusted so that $\omega_{\rm LC}=\omega_{\rm v0}$. The latter is true even if a dc magnetic field is applied along the disc plane. Under these circumstances, the vortex resonance frequency $\omega_{\rm v0}$ varies slightly (a few 10 MHz at maximum) as shown in Fig. 3 in the main text.  On the other side, $\kappa$ is fixed by design and typically takes values within a few kHz up to  $\sim 100$ MHz. The value of $\kappa$ will determine the maximum quality factor of the resonator $Q = \omega_{\rm LC} /\gamma_{\rm LC}$. In our simulations we use $\kappa/2\pi = 30$ MHz and {\color{black} $Q = 10^{3}$} which is feasible.  
%

The magnetic disc behaves as a resonator with characteristic frequency $\omega_{\rm v0}$ set by the aspect ratio and saturation magnetization $M_{\rm sat}$. Its loses are mainly given by the Gilbert damping $\alpha$ of the ferromagnetic material but they also depend on the particular excited mode. We determine both $\omega_{\rm v0}$ and $\gamma_{\rm v}$ by means of micromagnetic simulations with MUMAX3. Here, we use $M_{\rm sat}$, $\alpha$ and exchange stiffness $A$ according to literature values for each material, as summarized in the {\bf Methods} Section.

The coupling between the vortex and the superconducting resonator is calculated as described in Ref. \cite{MartinezPerez2018, MartinezPerez2019} and briefly summarized here. We first calculate the zero point current fluctuations flowing though the LC resonator:
\begin{eqnarray}
i_{\rm rms}=\omega_{\rm p}\sqrt{\frac{\hbar\pi}{4 Z_0}},
\label{irms}
\end{eqnarray}
with $Z_0$ the impedance. Next, we estimate the modulus of the zero point field fluctuations $B_{\rm rms,s}$ produced by $i_{\rm rms}$ at the vortex center position. $B_{\rm rms,s}$ depends on the thickness and width of the superconducting line. Finally, the coupling is calculated as 
\begin{equation}
g_{\rm LC}^2 = \frac{\kappa B_{\rm rms,s} \Delta M V}{4 \hbar}
\label{couplingdisc}
\end{equation}
where $V$ is the volume of the magnetic disc and $\Delta M$ is the amplitude of the volume averaged magnetization modulation when the vortex is excited with a varying magnetic field of amplitude $b_{\rm rms,s}$ at frequency $\omega_{\rm v0}$. The validity of this equation has been demonstrated by comparing numerical simulations based on \eqref{couplingdisc} with experimental coupling values from transmission and cavity measurements with Py nanomagnets \cite{Rincon2023}. 

In our simulations we use 3D-MLSI to estimate the distribution of supercurrents in the superconducting circuits and the resulting magnetic fields. Here, we have assumed superconducting Nb circuits with thickness $50$ nm and  width $100$ nm. The impedance of the LC resonator can be set by design and is assumed to be $Z_0=10$ $\Omega$.   By doing so, in the particular cases described in the main text we obtain $g_{\rm LC}/2\pi = 6.8$ MHz for the $400$ nm $\times 60$ nm Py disc and  $g_{\rm LC}/2\pi = 1.4$ MHz for the $200$ nm $\times 60$ nm YIG disc. 

Finally, we assume that  each spin is resonant with the vortex at frequency $\omega_{\rm v0} = \omega_{S} =\gamma_e B_{\rm tot}$ with $B_{\rm tot}=B_{\rm ap} + B_{\rm stray}$ the total static field at each spin position. $G$ is calculated as described in the main text.

\begin{acknowledgements}

This work is partly funded and supported by the European Research Council (ERC) under the European Union’s Horizon 2020 research and innovation programme (948986 QFaST), the spanish MCIN/AEI/10.13039/501100011033, the European Union
NextGenerationEU/PRTR and FEDER \emph{Una manera de hacer Europa} through projects EUR2019-103823, RTI2018-096075-B-C21, TED2021-131447B-C21 and PID2020-115221GB-C41, the CSIC program for the Spanish Recovery, Transformation and Resilience Plan funded by the Recovery and Resilience Facility of the European Union, established by the Regulation (EU) 2020/2094, the CSIC Research Platform on Quantum Technologies PTI-001, the Arag\'{o}n Regional Government through project QMAD (E09\_23R) and MCIN with funding from European Union NextGenerationEU (PRTR-C17.I1) promoted by the Government of Arag\'{o}n. Authors acknowledge fruitful discussions with Jes\'{u}s Mart\'{i}nez-Mart\'{i}nez, Alicia G\'{o}mez, Marina Calero and Sebasti\'{a}n Roca.
\end{acknowledgements}


\bibliography{sensing}

\end{document}